\documentstyle[aps,preprint,eqsecnum]{revtex}

\tightenlines

\begin{document}
\title{Baryon masses in a loop expansion with form factor}
\author{Phuoc Ha\thanks{Electronic address:
phuoc@theory1.physics.wisc.edu}
and  Loyal Durand\thanks{Electronic
address: ldurand@theory2.physics.wisc.edu}}
\address{
        Department of Physics, University of Wisconsin-Madison \\
             Madison, Wisconsin 53706, USA
        }
\date{\today}
\maketitle

\begin{abstract}
We show that the average multiplet masses in the baryon octet and
decuplet can be fitted with an average error of only $0.5\pm 0.3$ MeV
in a meson loop expansion with chiral SU(6) couplings, with the hadrons
treated as composite particles using a baryon-meson form factor. The
form factor suppresses unphysical short distance
effects and leads to a controllable expansion.
We find, in contrast to the results of standard chiral perturbation
theory, that pion loops are as
important as kaon or eta loops as would be expected when only
intermediate- and long-distance contributions are retained. We also
find that the contributions of decuplet intermediate states are important
in the calculation of the masses, and those states
must be included explicitly in a consistent theory. These results agree with
those of our recent loop-expansion analysis of the baryon magnetic moments.
We show, finally, that the parts of the loop contributions that change the
tree-level structure of the baryon masses are small, but largely account for
the violations of the baryon mass sum rules which hold at tree level.

\end{abstract}


\section{INTRODUCTION}

Baryon masses have been studied intensively in chiral perturbation
theory (ChPT). The results obtained in the standard approach using
dimensional regularization to control the divergences in the theory
are not completely satisfactory \cite{Jen,Beretal,Boretal}. The
chiral loop corrections evaluated in that regularization scheme are very
large, even of the order of the leading terms, and the convergence of the
chiral expansion is at best very slow \cite{Don2}.
In addition, the results are dominated by
the contributions from heavy mesons, while pion loops would be
expected to  give the similar contributions in the range of
distances or momenta in which the calculations are supposed to be reliable.
The same difficulties appear in other situations, for example, in the
calculation of the baryon magnetic moments in chiral perturbation
theory \cite{moments,Jenetal,Dai,Mei},
where the convergence and usefulness of the chiral expansion is again
questionable \cite{DandH}. These
shortcomings are due physically to the treatment of the
hadrons as point particles in the standard approach. Loop integrals then
involve high momenta and unphysical short-distance contributions, and
tend to be large and dominated by heavy mesons when calculated using
dimensional regularization. The extension of ChPT
to higher orders also requires the introduction
of new couplings in the effective field theory, and the theory
loses its predictive power.\footnote{ The chiral expansion
gives a complete parametrization of the static properties of the
baryons when carried to high enough order, but no dynamical information
unless the new couplings can be calculated in the underlying theory.
See, for example, the analysis
of the magnetic moment problem in \cite{DandH}.}

In this work, we reexamine the calculation of the baryon masses
using a loop expansion rather than an expansion in powers of the chiral
parameters. Our approach is motivated by the remarkable success of a
loop expansion for the baryon magnetic moments which starts from a QCD-based
quark model \cite{DandH1,DandH3}. That model was derived in a quenched
approximation to QCD following the Wilson-loop method of Brambilla
{\it et al.} \cite{brambilla}. The inclusion of loop corrections was
necessary to remove the quenching. The baryon Hamiltonian obtained
by Brambilla {\it et al.} has been used to fit the
baryon spectra with reasonable accuracy \cite{kogut,isgur}, but again in
the quenched approximation. We concentrate here on the calculation of the
one-loop corrections to the masses in the ground-state baryon octet
and decuplet. As in our earlier work on the moment problem \cite{DandH3},
we use chiral couplings for the low momentum interactions
of the mesons and baryons, but treat those hadrons as composite.
In particular, because of their extended structure, the baryons and mesons
cannot absorb high recoil momenta and remain in the same state,
so that high-momentum or short-distance effects are suppressed naturally
by wave function effects. We model these by introducing
a form factor at the baryon-meson vertex as in \cite{DandH1,DandH3,Ha}.
The divergences associated with point baryons are eliminated in this
approach, and the residual loop corrections are reduced in magnitude and
have  similar magnitudes for pion, kaon, and eta loops.

We include the decuplet baryon states explicitly in our calculations
as was done in \cite{Jen} and in our earlier work \cite{DandH3,Ha}.
We find that the contributions of octet and decuplet intermediate
states are of comparable importance for either the octet or decuplet
masses. This would be expected in a QM picture, since the
octet and decuplet baryons differ only in their spin structure.
We note also that the octet-decuplet mass splitting
is small on the scale of the relevant loop momenta, so an expansion
of the decuplet contributions in powers of the loop momentum
over the mass splitting as in some approaches to ChPT is poorly convergent
even when the calculation is restricted  to the octet masses.

The results of our calculation are in excellent agreement with
experiment. The loop corrections are typically about 35 percent of the
leading mass term, suggesting that the loop expansion converges
reasonably well. Moreover, the structure of the corrections is such
that they can largely be absorbed by adjustments of the input
parameters, leaving residual corrections to the baryon masses of less than
10 MeV. If we fit the octet and decuplet masses simultaneously
using the five parameters
that appear in the theory at tree level and the SU(6) couplings
used in our earlier work, we reproduce the average masses of the eight octet
and decuplet baryon multiplets with an average deviation of only
$0.5\pm 0.3$ MeV.
This is to be compared with the average deviation of $3.1\pm 0.3$ MeV in a fit
without loop corrections.
If we fix one of the parameters using a lowest order QM relation, the  average
deviation increases to 4.0 MeV, to be compared with a deviation
of 4.4 MeV in a corresponding fit without loop corrections. The real
improvement in the results becomes clear
when we analyze the structure of the loop corrections and the violations
of the mass relations that hold at
tree level. The parts of the loop corrections that cannot be described
within the tree-level structure are small, but largely account for the
violations of the mass sum rules.

An approach to loop corrections similar to that used here and in
our earlier work \cite{DandH1,DandH3,Ha} was recently introduced by
Donoghue and Holstein \cite{Don1} and Donoghue, Holstein, and Borasoy
\cite{Don2} in the context of ChPT.
Those authors show that the difficulties with dimensional
regularization in ChPT arise from unwanted
short-distance effects. They propose to eliminate those effects
by the use of a chiral cutoff in momentum space, while we have introduced
the cutoff as resulting physically from the composite structure of
the baryons \cite{DandH1,DandH3,Ha}. Their analysis shows that the effects
of uncertainty in the cutoff can largely be absorbed in a redefinition
of the input parameters such as occurs in our fitting procedure. However,
they do not include the baryon decuplet contributions in
their discussions of baryon masses and moments.

\section{MODEL AND THEORETICAL RESULTS}

\subsection{Heavy baryon couplings}

Heavy baryon perturbation theory (HBPT) was developed in
\cite{HBPT} and
extended to the chiral context in \cite{HBChPT}. It has been
used to study a number of hadronic processes at momentum transfers
much less than a typical baryon mass.
The key ideas in HBPT involve the replacement
of the momentum $p^\mu$ of a nearly on-shell baryon by its on-shell
momentum $m_Bv^\mu$ plus a small additional momentum $k^\mu$,
$p=m_Bv+k$, and the replacement of the baryon field operator $B(x)$
by an velocity-dependent operator $B_v(x)$ constructed to remove the
free momentum dependence in the Dirac equation,
$  B_v (x) = e^{i m_B {\not v} v^{\mu} x_{\mu}}B(x) \  $
\cite{HBPT}. In these expressions $m_B$ is the SU(3)-symmetric mass of
the baryon octet, $v^\mu$ is the on-shell four velocity
of the baryon, and it is assumed that $k\cdot v\ll m_B$.
Velocity-dependent  Rarita-Schwinger decuplet fields $T_v^\mu$ are
defined in the same manner,
with
$  T_v^\mu (x) = e^{i m_B {\not v} v^{\nu} x_{\nu}}T^\mu(x)$.
Note that we only extract the large octet-baryon mass $m_B$ in this
construction to avoid the appearance of phase factors in the
octet-decuplet
interactions defined below, and will treat the small decuplet-octet mass
difference $\delta m_{B}=m_T-m_B$ explicitly. The velocity-dependent
perturbation expansion involves modified Feynman rules and an expansion in
powers of $k/m_B$ \cite{HBPT,HBChPT}.

Because the low-momentum couplings of mesons to baryons appear to be well
described as derivative couplings with the standard chiral structure,
we will use those couplings in our analysis. However, we emphasize that we
will be making a loop expansion for the masses rather than the
expansion in powers of the momentum and the
symmetry breaking parameter $m_s$ characteristic of ChPT.
Higher-order effective couplings of ChPT will be implicit output
of our dynamical calculation.

The Lagrangian for the modified baryon fields depends on the usual
matrix of baryon fields, with $B$ replaced by $B_v$, and on the pseudoscalar
pion octet normalized as
\begin{equation}\label{eq:1}
\bbox{\phi} = {1 \over \sqrt{2}} \left(
\begin{array}{ccc}
{\pi^0\over\sqrt{2}} + {\eta\over \sqrt{6}} & \pi^+ & K^+ \\ \pi^- & -{\pi^0
\over \sqrt{2}} + {\eta \over \sqrt{6}} & K^0 \\ K^- & \overline K^0 & - {2\eta
\over \sqrt{6}}
\end{array} \right) \,.
\end{equation}
This couples to the baryon fields at low momenta through the
vector and axial vector currents defined by
\begin{eqnarray}
V_\mu = f^{-2} (\phi\partial_\mu\phi - \partial_\mu\phi \phi)+\cdots  \,, \ \
A_\mu&=& f^{-1}\partial_\mu \phi+\cdots \,, \ \
\end{eqnarray}
where $f \approx 93$ MeV is the meson decay constant. We will retain, as
indicated above, only leading term in the derivative expansion.
The lowest order Lagrangian for octet and decuplet baryons is then
\begin{eqnarray}\label{lag}
{\cal L}_v &=& i \ {\rm Tr}\ \bar B_v\ \left(v\cdot {\cal D}
\right)B_v
+ 2\ D\  {\rm Tr}\ \bar B_v\ S_v^\mu\ \{ A_\mu, B_v \}
+ 2\ F\ {\rm Tr}\  \bar B_v\ S_v^\mu\ [A_\mu, B_v]
\nonumber \\
&&-\ i\ \bar
T_v^{\mu}\ (v \cdot {\cal D}) \  T_{v \mu}
+ \delta m_B\ \bar T_v^{\mu}\ T_{v \mu}
+ {\cal C}\ \left(\bar T_v^{\mu}\ A_{\mu}\ B_v + \bar B_v\ A_{\mu}\
T_v^{\mu}\right){\phantom {f^2 \over 4}}
\nonumber\\
&& +\ 2\ {\cal H}\  \bar T_v^{\mu}\ S_{v \nu}\ A^{\nu}\  T_{v \mu}
+ {\rm Tr}\ \partial_\mu \phi \partial^\mu
\phi +\cdots \, \ \
\end{eqnarray}
where $\delta m_B$ is the decuplet-octet mass difference and
${\cal D_\mu}= \partial_\mu+[V_\mu,\,\cdot\, ]$ is the covariant
derivative. $B_v$ is now the matrix of octet baryon fields, and the
Rarita-Schwinger fields $T_v^\mu$ \cite{HBChPT} represent the decuplet
baryons. $D$, $F$, ${\cal C}$, and ${\cal H}$ are the strong interaction
coupling constants. The spin operator $S_v^{\mu}$ is defined in
Ref.\cite{HBChPT}. The ``mass term'' $\delta m_{B}\,\bar T_v^{\mu} T_{v \mu}$
for the decuplet fields will be absorbed in the following calculations
in the definition of the decuplet propagator. The propagator for the
fields $B_v$ involves no mass.

We will introduce SU(3) symmetry breaking into the
Lagrangian by including a quark mass
matrix $M={\rm diag}(m_u, m_d, m_s)$ in chiral combinations
with the velocity-dependent fields.
We will take $m_u=m_d=0$ in the present calculations as far as explicit
symmetry breaking is concerned, but will use the physical values of
the meson masses in evaluating the contributions of meson loop
corrections to the baryon masses. The tree-level masses splittings
of the baryons from $m_B$ are then defined to first order in $M$ by the
Lagrangian \cite{Jen}
\begin{eqnarray}
\label{eq:L^M}
{\cal L}_v^M &=& b_D \,{\rm Tr}\,\bar B_v \{ {\cal M},B_v \}
 + b_F \,{\rm Tr}\, \bar B_v [ {\cal M} , B_v ]
+ c \ \bar T_v^{\mu} {\cal M} T_{v \mu} \nonumber \\
&&+ \sigma \,{\rm Tr}\, M(\Sigma +
\Sigma^\dagger) \ {\rm Tr}\,\bar B_v  B_v
- \tilde\sigma \,{\rm Tr}\,M(\Sigma + \Sigma^\dagger)
\ \bar T_v^{\mu} T_{v \mu} \,,
\end{eqnarray}
where
\begin{equation}
{\cal M}=\xi^\dagger M \xi^\dagger + \xi M \xi,\qquad \xi=\exp (i\phi/f).
\end{equation}
If, as here,  we consider only baryon and not meson masses, the
terms with coefficients
$\sigma$ and $\tilde\sigma$ contribute mass terms of standard form
for the octet and decuplet baryons, and can be absorbed by
redefinitions
of $m_B$ and $m_T=m_B+\delta m_{B}$. We will follow this procedure. The
theory then involves five mass parameters at the tree level,
$ m_B$, $\delta m_{B}$, $m_s b_D$, $m_s b_F$, and $m_s c$.

It is straightforward to show that the structure at this level
encompasses that in the $L=0$ quark model taken to first order in
$m_s$, but the QM provides an extra first-order constraint
\begin{equation} \label{constrain}
c=-3(b_D+b_F)
\end{equation}
which connects the octet and decuplet masses.\footnote{See, for example, the
semirelativistic Hamiltonian for the
baryons given in \cite{brambilla,kogut,isgur}. Differentiation
of the Hamiltonian with respect
$m_s$ and use of the Feynman-Hellman theorem \cite{feynman}
gives first-order perturbations
of the baryon masses with precisely the structure given in Eq.\
(\ref{eq:L^M}), but with $\delta m_{B}$, $m_s b_D$, $m_s b_F$, and $m_s c$
expressed in terms of specific matrix elements. The relation for $c$
in terms of $b_D$ and $b_F$ follows.}
The existence of the
constraint is a consequence of the general spin structure of the
mass terms at that order, and the assumption that the
octet and decuplet baryons
would have the same internal structure up to spin for $m_s=0$. We will
consider both the constrained and unconstrained cases.

\subsection{Compositeness and the form factor}

In our previous work \cite{DandH1,DandH3} on a QCD-based quark
model with chiral couplings, we introduced a form factor characterizing
the structure of
baryons considered as composite particles and showed how to evaluate
the loop graphs with the form factor inserted at the
baryon-meson vertices.
We will follow the same approach here. In particular, we
introduce a simple form factor $F(k,v)$ at each meson-baryon vertex,
with
\begin{equation}
F(k,v)=\frac{\lambda^{2}}{\lambda^{2}+(k\cdot v)^2 -k^{2}}.
\end{equation}
Here $ k=(k_0, {\bf k})$ is the 4-momentum of meson and $\lambda$ is a
parameter characterizing a natural momentum scale for the meson-baryon wave
function. The form factor
reduces in the rest frame of the heavy baryon to a function of
${\bf k}^{2}$ only,
\begin{equation} \label{formf}
F(k,v) = { \lambda^2 \over {\lambda^2 + {\bf k}^2}} \ , \quad
v=(1,{\bf 0})\ ,
\end{equation}
and can be interpreted in terms of a meson-baryon wave function. It
also respects crossing symmetry for the meson line under the
substitution ${\bf k}\rightarrow -{\bf k}$.

The loop integrals involving this form factor can be evaluated by
making an appropriate $v$-dependent shift in the loop momentum
\cite{DandH3}.
Because of the structure of the vertices and the
relation $v\cdot S=0$, the result is equivalent for
some of the one-loop graphs
to that obtained with a covariant form factor or cutoff
$F(k)=\lambda^{2}/(\lambda^{2}-k^{2})$ \cite{Don2}, but the physical
content or motivation is different.

In  \cite{DandH3}, we found an excellent fit to the
baryon magnetic moments using a value $\lambda \approx 407$ MeV
in the form factor, but good fits could also be obtained for
somewhat different values in this general range.

\subsection{Expressions for the baryon masses}
\label{subsec:baryon_masses}
We will write the mass of baryon $i$ in the form
\begin{equation}
M_i = M_i^{(0)} + M_i^{(\delta m_{B}=0)} + M_i^{(\delta m_{B}\neq 0)}\ ,
\end{equation}
where the leading term $M_i^{(0)}$ includes the SU(3) symmetric mass
plus the tree-level mass splittings calculated from ${\cal L}_v^M$. The
terms in $M_i^{(\delta m_{B}=0)}$ are contributions from
the loop graphs in Figs.\ 1 and 2
that involve only octet or only decuplet baryons.
These contributions
are independent of the decuplet-octet mass difference
$\delta m_{B}$. The terms in $\mu_b^{(\delta m_{B}\neq 0)}$ come from
the loop graphs in Figs.\ 1b, 1d, 2c, and 2f
that involve both octet and decuplet baryons, and
depend explicitly on $\delta m_{B}$.

The calculations with the form factor $F(k,v)$ included
are straightforward using the methods in \cite{DandH3}, and we find
that\footnote{ For simplicity, we suppress all the subscripts and
superscripts used to label the loops.}
\begin{equation}
M_i^{(0)}= m^{(\pm)}+\alpha_i m_s  \ ,
\end{equation}
where $m^{(+)}= m_B$, $m^{(-)}=m_B+\delta m_{B}$,
\begin{eqnarray}
M_i^{(\delta m_{B}=0)} &=& \sum_{X=\pi,K,\eta} {1 \over {16\pi^2 f^2}}
\left \{ -\pi {\widetilde M}_X^3 \beta_i^{(X)}  \right.  \nonumber \\
       && +\left. 2m_s [ \, {\tilde \gamma}_i^{(X)} {\widetilde L}_0(
m_X, \lambda) +  ({\hat\gamma}_i^{(X)}-\lambda_i^{(X)}\alpha_i )
 L_0( m_X, \lambda) \, ] \right \} \ , \label{massde0}
\end{eqnarray}
and
\begin{eqnarray} \label{massde}
M_i^{(\delta m_{B}\neq 0)} &= & \sum_{X=\pi,K,\eta} {1 \over {16\pi^2
f^2}} \left \{ -\pi \beta_i^{'(X)} {\widetilde L}_2(m_X , \pm \delta
m_{B}, \lambda)  \right.  \nonumber \\
     && + \left. m_s \, ( \gamma_i^{'(X)}
  - \lambda_i^{'(X)}\alpha_i ) L_1(m_X , \pm \delta m_{B}, \lambda)
\right \} \ .
\end{eqnarray}
The upper and lower signs are to be used for external octet and decuplet
baryons, respectively. The functions ${\widetilde M}^3$,
${\widetilde L}_2(m , \pm \delta m_{B}, \lambda)$, and ${\widetilde L}_0(m ,
\lambda)$ are defined by
\begin{eqnarray}
\label{M3}
{\widetilde M}^3 &=& {\lambda^4 \over 2} {\lambda + 2m \over
(\lambda+m)^2} \ , \\
{\widetilde L}_2(m , \pm \delta m_{B}, \lambda) &=& {\widetilde M}^3 \mp
{\delta m_{B}\over \pi}
L_2(m , \pm \delta m_{B}, \lambda) \ , \\
{\widetilde L}_0(m , \lambda) &=&{\lambda^4 \over (\lambda^2-m^2)^2}
\Big [ \lambda^2-m^2
+ {m^2 \over \lambda} F_0(m, \lambda) \Big ] \ .
\end{eqnarray}
The remaining functions $F_0$, $L_0$, $L_1$, and
$L_2$  are given in \cite{DandH3} and \cite{Ha}.

The coupling coefficients $\alpha_i$ are identical to those in
\cite{Jen}. We list the remaining, mostly new,
coupling coefficients $\beta_i$, $\beta_i^{'}$,
${\hat\gamma}_i$, $\tilde \gamma_i$, $\gamma_i^{'}$,
$\lambda_i$, and $\lambda_i^{'}$ in Appendix A.
These reduce in combination to the coefficients given in \cite{Jen}
in the limit considered there.\footnote{For example, the sum
${\hat\gamma}_i + \tilde
\gamma_i+\gamma_i^{'}$ is denoted by $\gamma_i$ in  \cite{Jen}.}

To connect the various terms to the loop graphs in Figs.\
\ref{fig:bm1} and \ref{fig:bm2} we note that $\beta_i$,
$\beta_i^{'}$, $\tilde \gamma_i$, ${\hat\gamma}_i$, $\gamma_i^{'}$
are, respectively, the coefficients of the graphs 1a, 1b, 2a, 2b, and
2c for the octet baryons, and of the graphs
1c, 1d, 2d, 2e, and 2f for the decuplet baryons. The coefficients
$\lambda_i$ and $\lambda_i^{'}$ are associated respectively with the
wave function renormalizations that arise from
graphs 1a and 1b for external octet baryons, and
1c and 1d for external decuplet baryons.

\section{COMPARISON WITH EXPERIMENT}

\subsection{Fits to the data}

We will choose the strong interaction couplings $F$, $D$, $\cal C$, and
$\cal H$ to satisfy the SU(6) relations $ F=2D/3,\ {\cal C}=-2D$,
and ${\cal H}=-3D$ as in our previous work \cite{DandH,DandH3,Ha}.
For $F=0.5$, we then have $D=0.75$, ${\cal C}= -1.5$, and
${\cal H}= -2.15$, with
$F+D \approx |g_A/g_V|=1.26$. We use the values $f_\pi=93$ MeV and
$f_K=f_\eta=1.2 f_\pi$ \cite{Jenetal}.
While we fit the cutoff parameter $\lambda$, the fitted value of
is found to lie in a range consistent with that
in our earlier fits the baryon
magnetic moments \cite{DandH3} as it must. $\lambda$ is therefore not
a new or independent parameter.

\subsubsection{Unconstrained case}

In this case we have a total of five parameters
$m_B$, $\delta m_{B}$, $m_s b_D$, $m_s b_F$, and $m_s c$ to fit the
average masses of the eight SU(2) baryon multiplets.
An  equal-weight least-squares fit to the masses using only these
parameters and no loop corrections gives an average deviation of $3.1\pm 0.3$
MeV between the theoretical and experimental values of the masses for
$m_B=1191.6$ MeV, $\delta m_{B}=43.4$ MeV, $m_s b_D=30.1$ MeV,
$m_s b_F=-94.8$ MeV, and $m_s c=220.5$ MeV.
When the loop corrections are included in the fit
and the parameters readjusted,\footnote{We use the
values of the parameters $\delta m_{B}$, $m_s b_D$, $m_s b_F$,
and $m_s c$ obtained in the fit without loop corrections in evaluating
the latter. The mass insertions in the loops in Fig.\ \ref{fig:bm2} and
corresponding insertions on external lines correct for the deviations
of the baryon masses on the internal and external masses from the symmetric
values $m_B$ and $m_T=m_B+\delta m_B$. These are described quite well by the
tree-level fit. The greater part of the loop corrections preserves the
tree-level structure and can be absorbed in
the corresponding parameters as discussed in the following.}
the average deviation from experiment drops to
$0.5\pm 0.3$ MeV, a value significantly smaller than the mass
splittings within the multiplets.The results are summarized in Table
\ref{table:masses}. The best-fit
parameters are $ m_B=1590.9$ MeV, $\delta m_{B}=86.6$ MeV,
$m_s b_D=28.7$ MeV, $m_s b_F=-76.7$ MeV, and $m_s c=185.6$ MeV,
with $\lambda=441$ MeV. The results
are not especially sensitive to the value of the cutoff parameter.
Thus, a change to the value $\lambda=407$ MeV used in our analysis of
the baryon magnetic moments \cite{DandH3} changes the average
deviation of the theoretical masses from experiment of 0.7 MeV.
The fitted value of the parameter $m_sc=185.6$ MeV is reasonably close to,
but somewhat above the
value $-3(b_D+b_F)m_s=144$ MeV predicted by the QM relation in Eq.\ (\ref
{constrain}). The difference is consistent with the expected
enhancement of the attractive hyperfine interaction in the octet, and
the corresponding suppression of the repulsive interactions
in the decuplet \cite{isgur}, which gives $m_sc>-3m_s(b_D+b_F)$.

A detailed breakdown of the contributions of the
loop integrals to the fitted baryon masses is given
in Table \ref{table:loop_details}. All of the loop corrections lower
the baryon masses as expected in second order perturbation
theory. The corrections to the masses are substantial, but are still
small in comparison to the leading contributions, ranging from
31 to 37 percent of the tree-level value of $m_{B}$ and suggesting
reasonable convergence of the loop expansion. These results are
in marked contrast to the
results obtained in HBChPT for the point baryons, where the loop
contributions calculated using dimensional regularization are
comparable in size to the leading terms \cite{Don2}.

The results in Table \ref{table:loop_details} also show that the
contributions of decuplet intermediate states are very
important, of the same order as the contributions
from the octet. Because the mean tree-level
decuplet-octet mass splitting $\Delta m_{B}=\delta m_B+m_sc-\frac{5}{3}m_sb_F=
106$ MeV is small on
the scale of the typical momenta $k\approx\lambda=441$ MeV determined
by the form factor,
it is not justified to consider the decuplet as massive relative to the
octet, integrate out its explicit contributions, and attempt an
expansion of the decuplet
contributions in powers of $ k/\Delta m_{B}$ as in \cite{Mei}.

Table \ref{table:meson_loops} shows the contributions of loops involving
specific mesons to the total loop corrections.
We note that the pion loops are
very important, contrary to the result in HBChPT with
dimensional regularization. In that approach, the finite parts of the
loop integrals are proportional to the square or cube of the mass of the
meson in the loop. Pion loops are consequently strongly suppressed,
and were ignored, for example, in \cite{Jen}.
The effect of the form factor
is crucial here. The derivative couplings of mesons to baryons
emphasize high-momentum or short-distance contributions in the loop
integrals which are not given reliably by the theory.
These contributions are cut off by the form factors,
reflecting the compositeness and extended structures of the baryons
and mesons. With the form factors present, the momentum scale
in the loop integrations is set
primarily by the parameter $\lambda$,
and the integrals are dominated
by intermediate-range contributions.
The effects of the meson masses
on the loop integrals are suppressed accordingly.
For example, the ratios of the loop integrals
${\widetilde M}^3$, Eq.\ (\ref{M3}), corresponding to the diagram in
Fig.\ \ref{fig:bm1}a are $\pi : K : \eta = 1: 0.76 : 0.73$  for
$\lambda=441$ MeV, a result not very different
from the unit ratios that would hold for equal masses, but very different
from the ratios $\pi : K : \eta = 1 : 44 : 60$ in ChPT. Similar results
hold for the other integrals defined above.

The loop corrections shown in Tables \ref{table:loop_details}
and \ref{table:meson_loops} are much larger than the final changes in the
baryon masses obtained after adjusting the parameters
$m_B$, $\delta m_{B}$, $m_s b_D$, $m_s b_F$, and $m_s c$. This
suggests that the loop corrections have approximately the tree-level
structure, and can be parametrized up to small residual contributions
$\epsilon_i$ using the tree-level mass
relations with five parameters analogous to those above.
We note that the total loop corrections
introduce no new symmetry breaking when the meson masses in the loops
are taken as equal. The parts of the individual loop contributions that
do not have the tree-level structure therefore cancel, and the total
corrections can be absorbed
completely by a readjustment of the tree-level parameters. While this is no
longer possible for $M_{\pi}\not=M_{K}\not=M_{\eta}$, the effect of
unequal masses on the loop integrals
is suppressed because of the form factors as discussed above.
The  corrections therefore retain the tree-level structure
approximately, and the largest parts of the corrections can
be absorbed by adjusting the input parameters.\footnote{This
was shown analytically by Holstein {\it et al.\/} \cite{Don2} for the
purely octet case
in their discussion of the difference between a momentum cutoff and
dimensional regularization. It remains true in the
present calculation with decuplet contributions.}

In Table \ref{table:meson_loops} we show the residual contributions
$\epsilon_i$ obtained after subtracting
out tree-level parametrizations of the individual
and combined  meson loops.
The $\epsilon$'s for pion, kaon, and eta loops are much smaller than the
loop corrections themselves, with magnitudes less than 20 MeV.
The $\epsilon$'s for different mesons all have similar magnitudes, and
a varying pattern of signs. There is no suppression of pion loops.
The total residual $\epsilon_{\rm tot}$ would vanish identically for equal
meson masses, and the values given are significantly smaller than
the individual loop contributions because of the expected cancellations.

We also find that the sum of the fitted
values of $m_B$, $\delta m_{B}$, $m_s b_D$, $m_s b_F$, and $m_s c$
and the corresponding parameters for the loops reproduces the tree-level
parameters. The only new contributions of the loop corrections are therefore
in the $\epsilon$'s. We will return to this in Sec.\ \ref{sec:mass_relations}.

\subsubsection{Constrained case}

If the coupling $c$ is constrained by
the QM relation in Eq.\ (\ref{constrain}), we have four instead of
five parameters to fit average masses of the eight SU(2) baryon multiplets.
An equal-weight least-squares fit to the octet and decuplet masses
using only  those four parameters gives a average deviation from the
experiment of $4.4\pm 0.3$ MeV. With the loop corrections included,
we find a best
fit for $ m_B=1555.7$ MeV, $\delta m_{B}=100.8$ MeV, $m_s b_D=20.5$ MeV,
$m_s b_F=-81.4$ MeV, and $\lambda=432.2$ MeV, with an average deviation
from the experimental values of the multiplet masses of about $4.0\pm 0.3$ MeV
arising mainly from deviations in the $\Sigma$ and $\Xi$ masses in the octet.
The value $m_sc=-3m_s(b_D+b_F)=182.7$ MeV, which affects only the
decuplet masses in first order, is in excellent agreement with the
value of 185.6 MeV obtained in the unconstrained fit.
The constraint affects mostly the value the of the difference
$m_s(b_D-b_F)$, and worsens the fit to
the octet masses. The final results are summarized in Table
\ref{table:cmasses}.

Our overall conclusions with respect to the loop
corrections are the same as in the unconstrained
case. The corrections to the masses are substantial, but are still small in
comparison to the leading contributions, ranging from 30 to 35 percent of
the tree-level value of $m_{B}$. This again suggests
reasonable convergence of the loop expansion. The pion loops are
as important as kaon or eta loops, and the
contributions of decuplet intermediate states are very
important, of the same order as the contributions
from octet states.

\subsection{BARYON MASS RELATIONS}
\label{sec:mass_relations}

The unconstrained fit to the baryon masses given above is excellent.
However, the improvement
over the tree level fit is small if measured solely by the average
deviation between theory and experiment, and the overall fit
does not give as sensitive a test of the loop
corrections as might be expected. As emphasized above, the only parts of
the loop contributions that go beyond the tree-level structure are
those in the $\epsilon$'s in Table \ref{table:meson_loops}.
To get a stronger
test of the theory, we need to emphasize mass relations that are
independent of the tree-level parameters and of the large parts of the
loop corrections that have the same structure.
If the coupling $c$ is unconstrained, there are three such mass relations,
specifically the Gell-Mann--Okubo formula
\begin{equation} \label{go_rel}
{1 \over 4}( 3 M_\Lambda + M_\Sigma) - {1 \over 2}(M_N+M_\Xi)=0
\end{equation}
for the baryon octet, and two independent relations from
Gell-Mann's equal spacing rule for the baryon decuplet,
\begin{equation}
M_{\Sigma^*}-M_\Delta=M_{\Xi^*}-M_{\Sigma^*}=M_{\Omega}-M_{\Xi^*} \ .
\end{equation}
We will rearrange the latter to eliminate all tree-level parameters,
and will deal with the relations
\begin{eqnarray} \label{dec_rel}
&&M_{\Sigma^*}+M_{\Xi^*}-M_{\Delta}-M_{\Omega}=0\,,\nonumber\\
&&2M_{\Sigma^*}-M_\Delta-M_{\Xi^*}=0 \, , \\
&&2M_{\Xi^*}-M_{\Sigma^*}-M_{\Omega}=0 \, ,\nonumber
\end{eqnarray}
where the first is the sum of the remaining two.

In Table \ref{table:relations}, we show the violations of these mass
relations obtained from the loop graphs evaluated using the couplings
and the fitting parameters given in the unconstrained case
of the previous section. The violations of the Gell-Mann--Okubo
formula, Eq.(\ref{go_rel}), are quite small graph-by-graph, both absolutely
and on the scale of the loop contributions. In particular, it is not
necessary to have small loop contributions for the Gell-Mann--Okubo
relation to be well satisfied by the final octet-baryon masses.

The violations of the decuplet mass relations Eq.(\ref{dec_rel})
by the pion, kaon, and eta loop graphs are significant individually.
However, the total violations are small
because of cancellations. The theoretical results agree with the observed
violations in sign, and also agree reasonably well in magnitude given the
theoretical and experimental uncertainties. We conclude that the general
structure of the loop corrections is correct, but further contributions to
the final masses are clearly needed.

\section{CONCLUSIONS}

In this paper, we have considered the one-loop corrections to the baryon
masses in a loop expansion in which baryons are treated as composite
particles by introducing the form factor. Using the parameters
that appear in the theory at tree level and the SU(6) couplings
used in our earlier work, we fit the average masses of the eight octet
and decuplet baryon multiplets can be fitted with an average deviation
of $0.5\pm 0.3$ MeV. The fit is remarkably good given the absence of higher
order contributions and the experimental uncertainties.

As discussed above, the smallness of loop contributions relative
to the leading terms suggests that the perturbation series is under
control. We find also that
pion loops are quite important as would be expected in a calculation
dominated by intermediate range contributions. Both results are in sharp
contrast to the situation in conventional HBChPT with dimensional
regularization, where there is little indication that the perturbation
series converges, and
pion contributions are strongly suppressed.
In addition, we find that the decuplet states must be treated as light
on the scale of the octet, and included explicitly. In particular,
the contributions of decuplet intermediate states in loops
to the octet masses are as important
as the contributions of octet intermediate states.

Finally, we emphasize that the
sensitivity of the loop contributions to the masses of the mesons in
the loops is greatly reduced by the form factors.
As a result, the loop contributions
retain the basic structure of the tree-level masses,
and can largely be absorbed by readjustment of the tree-level
parameters. The overall
precision of the fit on the scale of the baryon masses is therefore
not a good test of the theory. However, the observed violations of
the Gell-Mann--Okubo and decuplet mass relations do provide tests.
We find that the corrected Gell-Mann--Okubo
relation is satisfied quite well. The pattern of
violations of the decuplet mass relations is also described
correctly, and the relations are satisfied
reasonably well numerically.

We believe that these results demonstrate the usefulness of treating baryons
as composite particles. By introducing the form factor, the
unphysical high-momentum effects that dominate conventional
calculations are suppressed, and a theory is obtained
that appears to describe low energy processes involving
baryons quite well. The dynamical problem that remains is one of
understanding the form factor at a deeper level.

\acknowledgments
This work was supported in part by the U.S. Department of Energy under
Grant No.\ DE-FG02-95ER40896. One of the authors (LD) would like to
thank the Aspen Center for Physics for its hospitality while the final
version of this paper was written.

\newpage
\appendix
\section{The coupling coefficients}
\setcounter{equation}{0}
In this appendix, we present the coupling coefficients explicitly.
For simplicity, the superscript $(X)$ is suppressed. Our coupling
coefficients $\alpha_i$ are identical to those in  \cite{Jen}.
There are the relations
between $\lambda_i$ and $\beta_i$ and also between $\lambda_i^{'}$
and
$\beta_i^{'}$
\begin{equation}
\lambda_i = {3 \over 2}\beta_i \ , \ \lambda_i^{'}= {3 \over
2}\beta_i^{'} \ .
\end{equation}

The coupling coefficients $\beta_i$ are
\begin{eqnarray} \label{betap}
 \beta_N^{(\pi)} &=& {3 \over 2}(D+F)^2 , \ \ \
 \beta_{\Sigma}^{(\pi)} = {2 \over 3} (D^2+6 F^2) , \nonumber \\
 \beta_{\Xi}^{(\pi)} &=& {3 \over 2}(D-F)^2 , \ \ \
 \beta_{\Lambda}^{(\pi)} = 2 D^2 , \nonumber \\
 \beta_{\Delta}^{(\pi)} &=& {25 \over 54}{\cal H}^2 , \ \ \
 \beta_{\Sigma^*}^{(\pi)} =  {20 \over 81}{\cal H}^2 ,  \\
 \beta_{\Xi^*}^{(\pi)} &=& {5 \over 54}{\cal H}^2 , \ \ \
 \beta_{\Omega}^{(\pi)} =  0 , \nonumber
\end{eqnarray}
for the pion loops,
\begin{eqnarray} \label{betak}
 \beta_N^{(K)} &=& {5 \over 3}D^2-2DF+3F^2 , \ \ \
 \beta_{\Sigma}^{(K)} = 2(D^2+F^2) , \nonumber \\
 \beta_{\Xi}^{(K)} &=& {5 \over 3}D^2+2DF+3F^2 , \ \ \
 \beta_{\Lambda}^{(K)} = {2 \over 3}D^2+6F^2 , \nonumber \\
 \beta_{\Delta}^{(K)} &=& {5 \over 27}{\cal H}^2 , \ \ \
 \beta_{\Sigma^*}^{(K)} =  {40 \over 81}{\cal H}^2 , \\
 \beta_{\Xi^*}^{(K)} &=& {5 \over 9}{\cal H}^2 , \ \ \
 \beta_{\Omega}^{(K)} = {10 \over 27}{\cal H}^2  , \nonumber
\end{eqnarray}
for the kaon loops, and
\begin{eqnarray} \label{betae}
 \beta_N^{(\eta)} &=& {1 \over 6}(D-3F)^2 , \ \ \
 \beta_{\Sigma}^{(\eta)} = {2 \over 3} D^2 , \nonumber \\
 \beta_{\Xi}^{(\eta)} &=& {1 \over 6}(D+3F)^2  , \ \ \
 \beta_{\Lambda}^{(\eta)} = {2 \over 3} D^2 , \nonumber \\
 \beta_{\Delta}^{(\eta)} &=& {5 \over 54}{\cal H}^2 , \ \ \
 \beta_{\Sigma^*}^{(\eta)} =  0 , \\
 \beta_{\Xi^*}^{(\eta)} &=& {5 \over 54}{\cal H}^2 , \ \ \
 \beta_{\Omega}^{(\eta)} = {10 \over 27}{\cal H}^2   , \nonumber
\end{eqnarray}
for the eta loops.

The coefficients $\beta_i^{'}$ are
\begin{eqnarray} \label{betpp}
 \beta_N^{'(\pi)} &=& {4 \over 3}{\cal C}^2 , \ \ \
 \beta_{\Sigma}^{'(\pi)} = {2 \over 9}{\cal C}^2  , \ \ \
 \beta_{\Xi}^{'(\pi)} = {{\cal C}^2 \over 3} , \ \ \
 \beta_{\Lambda}^{'(\pi)} = {\cal C}^2 , \nonumber \\
 \beta_{\Delta}^{'(\pi)} &=& {{\cal C}^2 \over 3} , \ \ \
 \beta_{\Sigma^*}^{'(\pi)} =  {5 \over 18}{\cal C}^2 , \ \ \
 \beta_{\Xi^*}^{'(\pi)} = {{\cal C}^2 \over 6} , \ \ \
 \beta_{\Omega}^{'(\pi)} =  0 ,
\end{eqnarray}
for the pion loops,
\begin{eqnarray} \label{betpk}
 \beta_N^{'(K)} &=& {{\cal C}^2 \over 3} , \ \ \
 \beta_{\Sigma}^{'(K)} = {10 \over 9}{\cal C}^2  , \ \ \
 \beta_{\Xi}^{'(K)} = {\cal C}^2 , \ \ \
 \beta_{\Lambda}^{'(K)} = {2 \over 3}{\cal C}^2 , \nonumber \\
 \beta_{\Delta}^{'(K)} &=& {{\cal C}^2 \over 3} , \ \ \
 \beta_{\Sigma^*}^{'(K)} =  {2 \over 9}{\cal C}^2 , \ \ \
 \beta_{\Xi^*}^{'(K)} = {{\cal C}^2 \over 3} , \ \ \
 \beta_{\Omega}^{'(K)} = {2 \over 3}{\cal C}^2  ,
\end{eqnarray}
for the kaon loops, and
\begin{eqnarray} \label{betpe}
 \beta_N^{'(\eta)} &=& 0 , \ \ \
 \beta_{\Sigma}^{'(\eta)} = {{\cal C}^2 \over 3} , \ \ \
 \beta_{\Xi}^{'(\eta)} = {{\cal C}^2 \over 3} , \ \ \
 \beta_{\Lambda}^{'(\eta)} = 0 , \nonumber \\
 \beta_{\Delta}^{'(\eta)} &=& 0 , \ \ \
 \beta_{\Sigma^*}^{'(\eta)} =  {{\cal C}^2 \over 6} ,\ \ \
 \beta_{\Xi^*}^{'(\eta)} = {{\cal C}^2 \over 6} , \ \ \
 \beta_{\Omega}^{'(\eta)} = 0 ,
\end{eqnarray}
for the eta loops.

The coefficients $\tilde \gamma_i$ are
\begin{eqnarray} \label{tgamk}
 \tilde \gamma_N^{(K)} &=& 3b_D-b_F+4\sigma , \
 \tilde \gamma_{\Sigma}^{(K)} = 2b_D+4\sigma  , \
 \tilde \gamma_{\Xi}^{(K)} = 3b_D+b_F+4\sigma , \
 \tilde \gamma_{\Lambda}^{(K)} = {10 \over 3}b_D+4\sigma , \nonumber
\\
 \tilde \gamma_{\Delta}^{(K)} &=& -c+4\tilde\sigma , \ \ \
 \tilde \gamma_{\Sigma^*}^{(K)} = -{4 \over 3}c+4\tilde\sigma ,  \ \
\
 \tilde \gamma_{\Xi^*}^{(K)} = -{5 \over 3}c+4\tilde\sigma , \ \ \
 \tilde \gamma_{\Omega}^{(K)} =  -2c+4\tilde\sigma  ,
\end{eqnarray}
for the kaon loops, and
\begin{eqnarray} \label{tgame}
 \tilde \gamma_N^{(\eta)} &=& {4 \over 3}(b_D-b_F+\sigma) , \
 \tilde \gamma_{\Sigma}^{(\eta)} = {4 \over 3}\sigma  , \
 \tilde \gamma_{\Xi}^{(\eta)} = {4 \over 3}(b_D+b_F+\sigma) , \
 \tilde \gamma_{\Lambda}^{(\eta)} = {4 \over 9}(4b_D+3\sigma) ,
\nonumber \\
 \tilde \gamma_{\Delta}^{(\eta)} &=& {4 \over 3}\tilde\sigma , \ \ \
 \tilde \gamma_{\Sigma^*}^{(\eta)} = {4 \over 9}(-c+3\tilde\sigma)
,  \ \ \
 \tilde \gamma_{\Xi^*}^{(\eta)} = {4 \over 9}(-2c+3\tilde\sigma) , \
\ \
 \tilde \gamma_{\Omega}^{(\eta)} =  {4 \over 3}(-c+\tilde\sigma)  ,
\end{eqnarray}
for the eta loops. The coefficients $\tilde \gamma_i$ vanish for
the pion loops.

The coefficients $\hat \gamma_i$ are
\begin{eqnarray} \label{hgamp}
 \hat \gamma_N^{(\pi)} &=& -{9 \over 2}(b_D-b_f)(D+F)^2
- 2\sigma\lambda_N^{(\pi)}, \ \ \
 \hat \gamma_{\Sigma}^{(\pi)} = -{8 \over 3}b_DD^2
- 2\sigma\lambda_{\Sigma}^{(\pi)} , \nonumber \\
 \hat \gamma_{\Xi}^{(\pi)} &=& -{9 \over 2}(b_D+b_f)(D-F)^2
- 2\sigma\lambda_{\Xi}^{(\pi)} , \ \ \
 \hat \gamma_{\Lambda}^{(\pi)} =
- 2\sigma\lambda_{\Lambda}^{(\pi)} , \nonumber \\
 \hat \gamma_{\Delta}^{(\pi)} &=&
- 2\tilde\sigma\lambda_{\Delta}^{(\pi)} , \ \ \
 \hat \gamma_{\Sigma^*}^{(\pi)} =  {20 \over 81}c{\cal H}^2
- 2\tilde\sigma\lambda_{\Sigma^*}^{(\pi)} , \\
 \hat \gamma_{\Xi^*}^{(\pi)} &=& {5 \over 27}c{\cal H}^2
- 2\tilde\sigma\lambda_{\Xi^*}^{(\pi)}, \ \ \
 \hat \gamma_{\Omega}^{(\pi)} =  -
2\tilde\sigma\lambda_{\Omega}^{(\pi)} , \nonumber
\end{eqnarray}
for the pion loops,
\begin{eqnarray} \label{hgamk}
 \hat \gamma_N^{(K)} &=& -{2 \over 3}b_D(D+3F)^2
- 2\sigma\lambda_N^{(K)}, \ \ \
 \hat \gamma_{\Sigma}^{(K)} = -6b_D(D^2+F^2)-12b_FDF
- 2\sigma\lambda_{\Sigma}^{(K)} , \nonumber \\
 \hat \gamma_{\Xi}^{(K)} &=& -{2 \over 3}b_D(D-3F)^2
- 2\sigma\lambda_{\Xi}^{(K)} , \ \ \
 \hat \gamma_{\Lambda}^{(K)} = -2b_D(D^2+9F^2)+12b_FDF
- 2\sigma\lambda_{\Lambda}^{(K)}, \nonumber \\
 \hat \gamma_{\Delta}^{(K)} &=& {5 \over 27}c{\cal H}^2
- 2\tilde\sigma\lambda_{\Delta}^{(K)} , \ \ \
 \hat \gamma_{\Sigma^*}^{(K)} =  {40 \over 81}c{\cal H}^2
- 2\tilde\sigma\lambda_{\Sigma^*}^{(K)} , \\
 \hat \gamma_{\Xi^*}^{(K)} &=& {5 \over 27}c{\cal H}^2
- 2\tilde\sigma\lambda_{\Xi^*}^{(K)} , \ \ \
 \hat \gamma_{\Omega}^{(K)} = {20 \over 27}c{\cal H}^2
- 2\tilde\sigma\lambda_{\Omega}^{(K)} , \nonumber
\end{eqnarray}
for the kaon loops, and
\begin{eqnarray} \label{hgame}
 \hat \gamma_N^{(\eta)} &=& -{1 \over 2}(b_D-b_F)(D-3F)^2
- 2\sigma\lambda_N^{(\eta)} , \ \ \
 \hat \gamma_{\Sigma}^{(\eta)} =
- 2\sigma\lambda_{\Sigma}^{(\eta)} , \nonumber \\
 \hat \gamma_{\Xi}^{(\eta)} &=& -{1 \over 2}(b_D+b_F)(D+3F)^2
- 2\sigma\lambda_{\Xi}^{(\eta)}  , \ \ \
 \hat \gamma_{\Lambda}^{(\eta)} = -{8 \over 3}b_D D^2
- 2\sigma\lambda_{\Lambda}^{(\eta)}, \nonumber \\
 \hat \gamma_{\Delta}^{(\eta)} &=&
- 2\tilde\sigma\lambda_{\Delta}^{(\eta)} , \ \ \
 \hat \gamma_{\Sigma^*}^{(\eta)} =
- 2\tilde\sigma\lambda_{\Sigma^*}^{(\eta)} , \\
 \hat \gamma_{\Xi^*}^{(\eta)} &=& {5 \over 27}c{\cal H}^2
- 2\tilde\sigma\lambda_{\Xi^*}^{(\eta)}, \ \ \
 \hat \gamma_{\Omega}^{(\eta)} = {10 \over 9}c{\cal H}^2
- 2\tilde\sigma\lambda_{\Omega}^{(\eta)}  , \nonumber
\end{eqnarray}
for the eta loops.

Finally, the coefficients $\gamma_i^{'}$ are
\begin{eqnarray} \label{gampp}
 \gamma_N^{'(\pi)} &=&
- 2\tilde\sigma\lambda_N^{'(\pi)} , \ \ \
 \gamma_{\Sigma}^{'(\pi)} = {2 \over 9}c \, {\cal C}^2
- 2\tilde\sigma\lambda_\Sigma^{'(\pi)} , \nonumber \\
 \gamma_{\Xi}^{'(\pi)} &=& {2 \over 3}c \, {\cal C}^2
- 2\tilde\sigma\lambda_\Xi^{'(\pi)}, \ \ \
 \gamma_{\Lambda}^{'(\pi)} = c \, {\cal C}^2
- 2\tilde\sigma\lambda_\Lambda^{'(\pi)}, \nonumber \\
 \gamma_{\Delta}^{'(\pi)} &=& (b_F-b_D){\cal C}^2
- 2\sigma\lambda_{\Delta}^{'(\pi)}, \ \ \
 \gamma_{\Sigma^*}^{'(\pi)} =  -{2 \over 3}b_D{\cal C}^2
- 2\sigma\lambda_{\Sigma^*}^{'(\pi)} , \nonumber \\
 \gamma_{\Xi^*}^{'(\pi)} &=& -{1 \over 2}(b_D+b_F){\cal C}^2
- 2\sigma\lambda_{\Xi^*}^{'(\pi)} , \ \ \
 \gamma_{\Omega}^{'(\pi)} =
- 2\sigma\lambda_{\Omega}^{'(\pi)} ,
\end{eqnarray}
for the pion loops,
\begin{eqnarray} \label{gampk}
 \gamma_N^{'(K)} &=& {1 \over 3} c \, {\cal C}^2
- 2\tilde\sigma\lambda_N^{'(K)}, \ \ \
 \gamma_{\Sigma}^{'(K)} = {4 \over 9}c \, {\cal C}^2
- 2\tilde\sigma\lambda_\Sigma^{'(K)} , \nonumber \\
 \gamma_{\Xi}^{'(K)} &=& {7 \over 3} c \, {\cal C}^2
- 2\tilde\sigma\lambda_\Xi^{'(K)}, \ \ \
 \gamma_{\Lambda}^{'(K)} = {4 \over 3}c \, {\cal C}^2
- 2\tilde\sigma\lambda_\Lambda^{'(K)} , \nonumber \\
 \gamma_{\Delta}^{'(K)} &=&
- 2\sigma\lambda_{\Delta}^{'(K)}, \ \ \
 \gamma_{\Sigma^*}^{'(K)} = - {2 \over 3}b_D{\cal C}^2
- 2\sigma\lambda_{\Sigma^*}^{'(K)} , \nonumber \\
 \gamma_{\Xi^*}^{'(K)} &=& - {2 \over 3}b_D{\cal C}^2
- 2\sigma\lambda_{\Xi^*}^{'(K)} , \ \ \
 \gamma_{\Omega}^{'(K)} = -2(b_D+b_F){\cal C}^2
- 2\sigma\lambda_{\Omega}^{'(K)} ,
\end{eqnarray}
for the kaon loops, and
\begin{eqnarray} \label{gampe}
 \gamma_N^{'(\eta)} &=&
- 2\tilde\sigma\lambda_N^{'(\eta)} , \ \ \
 \gamma_{\Sigma}^{'(\eta)} = {1 \over 3}c \, {\cal C}^2
- 2\tilde\sigma\lambda_\Sigma^{'(\eta)} , \nonumber \\
 \gamma_{\Xi}^{'(\eta)} &=& {2 \over 3}c \, {\cal C}^2
- 2\tilde\sigma\lambda_\Xi^{'(\eta)}, \ \ \
 \gamma_{\Lambda}^{'(\eta)} =
- 2\tilde\sigma\lambda_\Lambda^{'(\eta)}, \nonumber \\
 \gamma_{\Delta}^{'(\eta)} &=&
- 2\sigma\lambda_{\Delta}^{'(\eta)} , \ \ \
 \gamma_{\Sigma^*}^{'(\eta)} =
- 2\sigma\lambda_{\Sigma^*}^{'(\eta)} , \nonumber \\
 \gamma_{\Xi^*}^{'(\eta)} &=& -{1 \over 2}(b_D+b_F){\cal C}^2
- 2\sigma\lambda_{\Xi^*}^{'(\eta)}, \ \ \
 \gamma_{\Omega}^{'(\eta)} =
- 2\sigma\lambda_{\Omega}^{'(\eta)} ,
\end{eqnarray}
for the eta loops.

\newpage

\begin{figure}
\caption{Diagrams that give rise to non-analytic
$m_s^{3/2}$ corrections to the baryon massess in conventional
HBChPT. The dashed lines denote the mesons, the single and double solid
lines
denote octet and decuplet baryons, respectively. A heavy dot with a
meson
line represents an insertion of the form factor $F(k,v)$
in Eq.(\protect\ref{formf}),
with meson momentum $k$. }
\label{fig:bm1}
\end{figure}

\begin{figure}
\caption{Diagrams that give rise to non-analytic
$m_s^2 \ln m_s$ corrections to the baryon masses in
conventional HBChPT. Two short straight lines denote an insertion of
the tree level mass terms $b_D$, $b_F$, and $\sigma$ for the
baryon octet, and the terms $c$ and $\tilde\sigma$ for the baryon decuplet.}
\label{fig:bm2}
\end{figure}

\newpage

\begin{table}
\caption{Baryon masses in MeV for the unconstrained case. \ Here
$\Delta M_i=M_i^{({\rm theory})}- M_i^{({\rm exp})}$. The average
deviation $|\overline{\Delta M}_i|=0.5\pm 0.3$ MeV.}
\label{table:masses}
\begin{tabular}{lrrrrrrrr}
      & $N$ & $\Lambda$ & $\Sigma$ & $\Xi$ & $\Delta$ & $\Sigma^*$ &
    $\Xi^*$ & $\Omega$ \\
    \hline
    Theory & 939.2 & 1115.3 & 1193.0 & 1318.4 & 1231.5 & 1385.8 &
1532.6 & 1672.6 \\
    Exp. & 938.9 & 1115.7 & 1193.1 & $1318.1 \pm 0.3$ & $1232.0 \pm 2.0$
    & $1384.6 \pm 0.4$ & $1533.4 \pm 0.3$ & $1672.5 \pm 0.3$ \\
    $\Delta M_i$ & 0.3 & $-$0.4 & $-$0.1 & $0.3 \pm 0.3$ & $-0.5 \pm 2.0$
& $1.2 \pm 0.4$ & $-0.8 \pm 0.3$ & $0.1 \pm 0.3$
  \end{tabular}
\end{table}

\begin{table}
\caption{Detailed breakdown of the contributions of the loop integrals
to the masses of the octet and decuplet baryons in MeV for the unconstrained
case. The contributions are evaluated using the value $\lambda =441$
MeV for the cutoff parameter,
and the couplings $F=0.5$, $D=0.75$, $ {\cal C} =-1.5$,
and $ {\cal H} =-2.15$. A best fit is obtained for $ m_B=1590.9$ MeV,
$m_s b_D=28.7$ MeV, $m_s b_F=-76.7$ MeV, $m_s c=185.6$ MeV, and
$\delta m_{B}=86.6$ MeV. Contributions from the various loop graphs are labeled
using the corresponding coupling coefficients defined in
Sec.\ \protect\ref{subsec:baryon_masses}.
}
\label{table:loop_details}
\begin{tabular}{cddddddd}
 Baryon & $\alpha$ & $\beta$ & $(\hat\gamma+\tilde\gamma-\lambda\alpha)$ &
$\beta^{'}$  & $(\gamma^{'}-\lambda^{'}\alpha)$ & Loops & $M_i$  \\
\hline
$N$ & 1380.1 & $-$268.5 & 49.1 & $-$301.5 & 80.0 & $-$440.9 & 939.2 \\
$\Lambda$ & 1514.3 & $-$214.7 & 20.2 & $-$270.9 & 66.3 & $-$399.1 &
1115.3 \\
$\Sigma$ & 1590.9 & $-$225.6 & 10.9 & $-$198.2 & 15.0 & $-$397.9 &
1193.0 \\
$\Xi$ & 1686.9 & $-$168.7 & $-$22.4 & $-$208.4 & 31.0 & $-$368.5 &
1318.4 \\
$\Delta$ & 1677.5 & $-$286.3 & $-$22.8 & $-$111.8 & $-$25.2 &
$-$446.0 & 1231.5 \\
$\Sigma^*$ & 1801.2 & $-$239.3 & $-$48.6 & $-$105.1 & $-$22.4 &
$-$415.5 & 1385.8 \\
$\Xi^*$ & 1924.9 & $-$204.2 & $-$74.5 & $-$93.3 & $-$20.3 & $-$392.3
& 1532.6 \\
$\Omega$ & 2048.7 & $-$181.1 & $-$100.4 & $-$76.3 & $-$18.3 & $-$376.0
& 1672.6 \\
\end{tabular}
\end{table}

\begin{table}
\caption{The contributions in MeV of the loop integrals involving pions,
 kaons, and etas to the baryon masses for the parameters given in Table II.
The quantities $\epsilon_i$ are the residual contributions of the loops
after the parts that can be described by the tree-level structure are
subtracted.}
\label{table:meson_loops}
\begin{tabular}{cddddddd}
Baryon & $\pi$ loops& $\epsilon_\pi$ & $K$ loops & $\epsilon_K$ & $\eta$
loops & $\epsilon_\eta$ & $\epsilon_{\rm tot}$\\
\hline
$N$ & $-$413.3 & 5.2 & $-$43.1 & $-$8.0 & 15.5 & 5.4 & 2.62 \\
$\Lambda$ & $-$245.9 & $-$7.9 & $-$141.8 & 12.0 & $-$11.3 & $-$8.0 & $-$3.93\\
$\Sigma$ & $-$168.5 & $-$2.6 & $-$180.4 & 4.0 & $-$49.0 & $-$2.7 & $-$1.31 \\
$\Xi$ & $-$63.5 & 5.3 & $-$223.8 & $-$8.0 & $-$81.1 & 5.4 & 2.62 \\
$\Delta$ & $-$316.8 & 8.6 & $-$107.0 & $-$17.0 & $-$22.2 & 12.0 & 3.57\\
$\Sigma^{*}$ & $-$194.0 & $-$8.7 & $-$189.4 & 17.1 & $-$32.0 & $-$12.0 &
$-$3.61\\
$\Xi^{*}$ & $-$88.7 & $-$8.4 & $-$237.4 & 16.8 & $-$66.2 & $-$11.9 & $-$3.49\\
$\Omega$ & 0.0 & 8.5 & $-$252.0 & $-$16.9 & $-$124.0 & 12.0 & 3.53\\
\end{tabular}
\end{table}

\begin{table}
\caption{Baryon masses in MeV for the constrained case. \ Here
$\Delta M_i=M_i^{({\rm theory})}- M_i^{({\rm exp})}$. The average
deviation $|\overline{\Delta M}_i|=4.0\pm 0.3$ MeV.}
\label{table:cmasses}
\begin{tabular}{lcccccccc}
      & $N$ & $\Lambda$ & $\Sigma$ & $\Xi$ & $\Delta$ & $\Sigma^*$ &
    $\Xi^*$ & $\Omega$ \\
    \hline
    Theory & 936.6 & 1119.2 & 1184.2 & 1325.9 & 1234.5 & 1386.7 &
1531.7 & 1669.6 \\
    Exp. & 938.9 & 1115.7 & 1193.1 & $1318.1 \pm 0.3 $ & $1232.0 \pm 2.0$
    & $1384.6 \pm 0.4$ & $1533.4 \pm 0.3$ & $1672.5 \pm 0.3$\\
    $\Delta M_i$ & $-$2.3 & 3.5 & $-$8.9 & $7.8 \pm 0.3$ & $2.5 \pm 2.0$
& $2.1 \pm 0.4$ & $-1.7 \pm 0.3$ & $-2.9 \pm 0.3$
  \end{tabular}
\end{table}

\begin{table}
\caption{Violations of the baryon mass relations in MeV for the unconstrained
case. Violations from the various loop graphs are labeled using the
corresponding coupling coefficients.
The tadpole graphs satify the mass relations.}
\label{table:relations}
\begin{tabular}{ldddddc}
     Relation & $\beta$ &$\beta^{'}$ & $(\hat\gamma-\lambda\alpha)$ &
$(\gamma^{'}-\lambda^{'}\alpha)$ & Total & Experiment  \\
\hline
Gell-Mann--Okubo & 1.2 & 2.2 & 4.5 & $-$2.0 & 5.9 & $6.6\pm 0.2$ \\
$M_{\Sigma^*}+M_{\Xi^*}-M_\Delta-M_{\Omega}$ & 23.8 & $-$10.4 & 0.0 &
0.8 & 14.2 & $13.5\pm 2.1$\\
$2M_{\Sigma^*}-M_\Delta-M_{\Xi^*}$ & 11.9 & $-$5.2 & 0.0
& 0.7 & 7.4 & $3.8\pm 2.1$ \\
$2M_{\Xi^*}-M_{\Sigma^*}-M_{\Omega}$ & 11.9 & $-$5.2 & 0.0
& 0.1 & 6.8 & $9.7\pm 0.7$ \\
\end{tabular}
\end{table}

\newpage

\input epsf
\begin{center}
\epsfbox{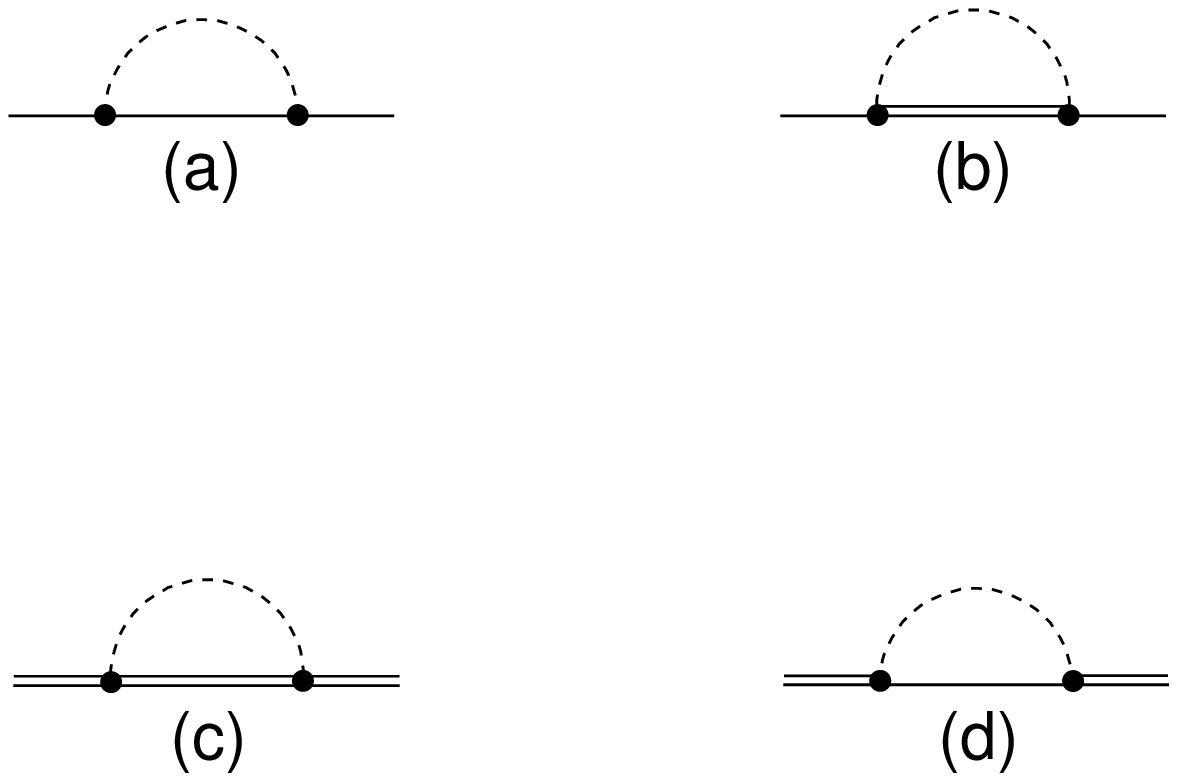}
\end{center}

\newpage

\input epsf
\begin{center}
\epsfbox{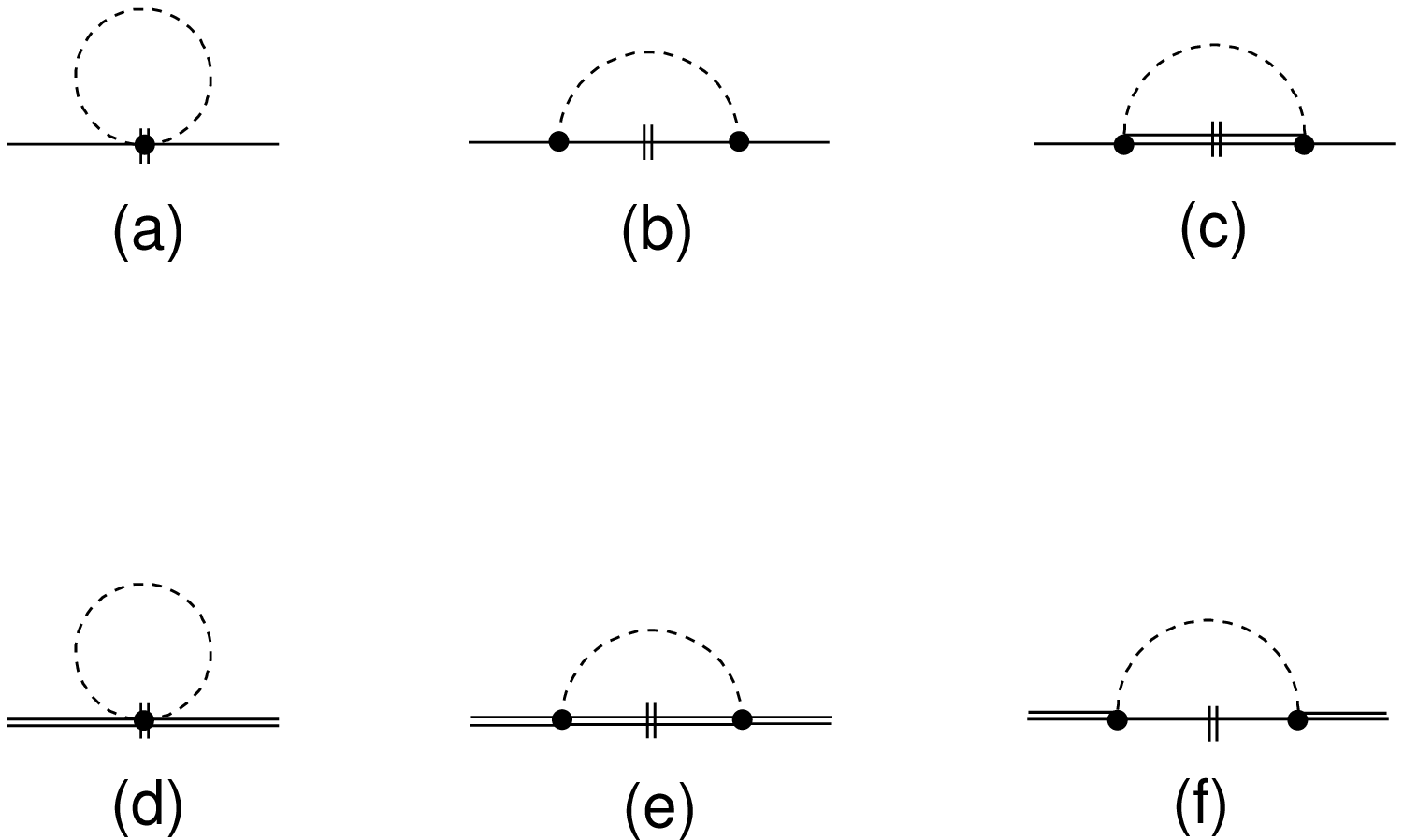}
\end{center}


\begin{references}
%
\bibitem{Jen}
E. Jenkins, Nucl.\ Phys.\  {\bf B368}, 190 (1992).
%
\bibitem{Beretal}
V. Bernard, N. Kaiser and U.G. Meissner, Z.\ Phys.\ C {\bf 60}, 111
(1993).
%
\bibitem{Boretal}
B. Borasoy and U.G. Meissner, Phys.\ Lett.\ B {\bf 365}, 285 (1996);
Ann.\ Phys.\ (NY) {\bf 254}, 192 (1997).
%
\bibitem{Don2}
J. Donoghue, B.R. Holstein, and B. Borasoy, hep-ph/9804281.
%
\bibitem{moments}
D.G. Caldi and H. Pagels, Phys.\ Rev.\ D {\bf 10}, 3739 (1974);
J. Gasser, M.
Sainio, and A. Svarc, Nucl.\ Phys.\ {\bf B307}, 779 (1988); A.
Krause, Helv.\
Phys.\ Acta {\bf 63}, 3 (1990).
%
\bibitem{Jenetal}
E. Jenkins, M. Luke, A.V. Manohar and M.
Savage, Phys.\ Lett.\ B {\bf 302},
482 (1993);  (E) {\it ibid.} {\bf 388}, 866 (1996).
%
\bibitem{Dai}
J. Dai, R. Dashen, E. Jenkins, and A.V. Manohar, Phys.\ Rev.\ D {\bf 53},
273 (1996).
%
\bibitem{Mei}
Ulf-G. Mei{\ss}ner and S. Steininger, Nucl.\ Phys.\ {\bf B499}, 349
(1997).
%
\bibitem{DandH}
L. Durand and P. Ha, Phys.\ Rev.\ D {\bf 58}, 013010
(1998).
%
\bibitem{DandH1}
L. Durand and P. Ha, {\it Quark Confinement
and Hadron Spectrum II}, edited by N. Brambilla and G.M. Prosperi (World
Scientific, Singapore, 1996).
%
\bibitem{DandH3}
P. Ha and L. Durand, Phys.\ Rev.\ D {\bf 58}, 093008 (1998).
%
\bibitem{brambilla}
N. Brambilla, P. Consoli, and G.M. Prosperi,
Phys.\ Rev.\ D {\bf 50}, 5878 (1994).
%
\bibitem{kogut}
J. Carlson, J. Kogut, and V.R. Pandharipande, Phys.\ Rev.\ D
{\bf 27}, 233 (1983); {\bf 28}, 2807 (1983).
%
\bibitem{isgur}
S. Capstick and N. Isgur, Phys.\ Rev.\ D {\bf 34}, 2809 (1986).
%
\bibitem{Ha}
P. Ha, Phys.\ Rev.\ D {\bf 58}, 113003 (1998).
%
\bibitem{Don1}
J. Donoghue and B.R. Holstein, hep-ph/9803312.
%
\bibitem{HBPT}
H. Georgi, Phys.\ Lett.\ B {\bf 240}, 447 (1990).
%
\bibitem{HBChPT}
E. Jenkins and A.V. Manohar, Phys.\ Lett.\ B {\bf 255}, 558 (1991);
{\it ibid}. {\bf 259}, 353 (1991);  Report No. UCSD/PTH 91-30.
%
\bibitem{feynman}
R.P. Feynman, Phys.\ Rev.\ {\bf 56 }, 340 (1939).
\end{references}
\end{document}